# Electronic circuit realization of the logistic map

MADHEKAR SUNEEL
PGAD (DRDO, Ministry of Defence, Government of India), DRDL Complex, Kanchanbagh, Hyderabad 500 058, India
e-mail: suneel@ieee.org

**Abstract.** An electronic circuit realization of the logistic difference equation is presented using analog electronics. The behavior of the realized system is evaluated against computer simulations of the same. The circuit is found to exhibit the entire range of dynamics of the logistic equation: fixed points, periodicity, period doubling, chaos and intermittency. Quantitative measurements of the dynamics of the realized system are presented and are found to be in good agreement with the theoretical values. Some possible applications of such a realization are briefly discussed.

**Keywords.** Nonlinear dynamics; dynamical systems; chaos; analog computation.

## 1. Introduction

Since the discovery of deterministic chaos in the mid 1960s, the field of non-linear dynamical systems has attracted the attention of researchers worldwide. Non-linear dynamical systems have found applications in areas as diverse as biophysics, meteorology, hydrodynamics, chemical engineering, optics, cryptology and communications. There has been considerable interest in finding simple circuits that exhibit non-linear phenomena such as bifurcations and chaos (Hasler 1987, Matsumoto 1987, Sprott & Linz 2000). Reports about realizations of mathematically known dynamical systems have been comparatively few (Kiers & Schmidt 2004). In this paper, an attempt is made to accurately realize a well-known dynamical system, namely the logistic difference equation, as an electronic circuit. It is found that such circuits cannot only be simple to implement but are also good engineering models of the respective mathematical systems. Murali et al (Murali et al 2003) have proposed a similar though not identical implementation of the logistic equation for the purpose of implementing a NOR gate. The circuit implemented the logistic map for a single value (4.0) of the control parameter. In this paper, however, a generalized circuit is designed that allows the variation of the control parameter over the entire range of 0 to 4. The variation of the control parameter allows the bifurcation diagram of the realized system to be experimentally obtained. Quantitative measurements of accuracy of the realized system are obtained by comparing the values of the control parameter at some key events of this bifurcation diagram with their computer-simulation counterparts. It is also interesting to note that Murakoshi et al (Murakoshi et al 2000) have demonstrated an approach to generate Pulse Width Modulated (PWM) and Pulse Position Modulated (PPM) signals of the logistic map. For this, they have used fixed precision digital hardware, D/A converters and analog filters along with PWM voltage conversion circuitry.

Discrete-time dynamical systems are a particular type of non-linear dynamical systems generally described as an iterative map $f : \Re^n \to \Re^n$ by the state equation

$$x_{k+1} = f(x_k), \quad k = 0, 1, 2, \cdots \tag{1}$$

where $n$ is the dimensionality of the state-space, $x_k \in \Re^n$ is the state of the system at time $k$. $x_{k+1}$ denotes the next state. Therefore, $k$ denotes the discrete time. Repeated iteration of $f$ gives rise to a sequence of points $\{x_k\}_{k=0}^{\infty}$, known as an *orbit*. Clearly, (1) is a difference equation.

The logistic map is a dynamical system described by the state equation

$$x_{k+1} = \lambda x_k (1 - x_k), \quad k = 0, 1, 2, \cdots \qquad (2)$$

where $x_k \in (0,1)$ and $\lambda \in (0,4)$. $\lambda$ is known as the *control parameter*.

## 2. Dynamics of the Equation

In spite of the apparent simplicity of (2) it is known to exhibit dynamics that is extremely complicated (May 1976). A very brief introduction to the dynamics is presented here. For more details, the reader is referred to any of the specialized works on the subject (Baker & Gollub 1990, Kumar 1996). The behavior of (2) is sensitive to the value of $\lambda$. For $0 \leq \lambda \leq 1$, the system always converges to the steady state $x = 0$. However, from $\lambda = 1$ the system ceases to converge to 0 and instead, converges to a different fixed point. In the interval $1 \leq \lambda \leq 3$, each value of $\lambda$ has a different steady-state value of $x$. At $\lambda = 3$, the system ceases to settle to any fixed point. Instead, it begins to oscillate between two states – a phenomenon known as *bifurcation*. As $\lambda$ is further increased in the interval $3 \leq \lambda \leq 3.57$, the system undergoes a series of bifurcations leading to long-term oscillatory behavior between four states, eight states, sixteen states and so on – a phenomenon known as *period doubling*. In the interval $3.58 \leq \lambda \leq 4$, the behavior of the system is non-convergent and non-periodic. The state-variable fills out continuous intervals instead of taking on values from a finite set. Such behavior is known as *chaos* and the system is said to be in a *chaotic regime*. However, in the interval $3.58 \leq \lambda \leq 4$ itself, the regions of chaos are found to be interspersed with brief windows of periodicity; a phenomenon known as *intermittency*. The complex behavior of the system is best understood by studying a graph plotted between $\lambda$ and the long-term value of $x$. Such a diagram, known as a *bifurcation diagram*, is shown in figure 1.

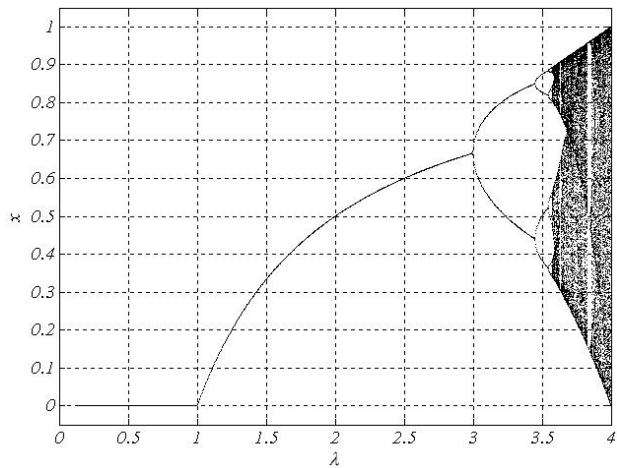

Figure 1. Bifurcation Diagram of the Logistic Difference Equation

One of the most interesting characteristics of chaotic dynamics is the extreme sensitivity to initial conditions. A consequence of this sensitivity is that two chaotic orbits, originating from two points arbitrarily close in state-space, rapidly diverge and become practically uncorrelated after the passage of a fair amount of time. Physical measurement of the initial state will always be associated with some amount of error and the sensitivity to initial conditions ensures that such errors make it impossible to predict chaotic orbits in the long run. Also, two identical systems do not give rise to identical orbits because there is bound to be at least a minute natural difference in the initial state that is sufficient to make the orbits divergent and uncorrelated.

It should be noted that chaotic orbits obtained using numerical methods and computer simulations actually turn out to be periodic because of the finite precision of the machines and the resulting approximations that take place at each step. Also, because the computed solutions do not depend on natural variation of conditions, the orbits are deterministic and identical algorithms generate identical orbits when provided with identical initial conditions.

### 3. Circuit Scheme

To realize (2), the presence of a block with the input-output relation

$$c(a,b) = \frac{ab}{10} \quad (3)$$

is assumed. Here $a, b$ are input voltages and $c$ is the output voltage. This is the default input-output characteristic of commercially available analog multiplier ICs. In (2), it can be observed that the values of the state variable lie in the interval $(0,1)$. However, most analog multiplier ICs have the operational input voltage range of –10 V to 10 V. Therefore, to avoid errors at small signal voltages, the state variable can safely be scaled by a factor of 10. For the same reason, the value of the control parameter $\lambda$ also can be scaled by a factor of $2.5$. This amounts to making the substitutions

$$\left. \begin{array}{l} X = 10x \\ \Lambda = 2.5\lambda \end{array} \right\} \quad (4)$$

These substitutions in (2), after simplification, yield

$$X_{k+1} = \frac{\Lambda X_k (10 - X_k)}{25}. \quad (5)$$

The circuit scheme of figure 2 is proposed to realize (5). In the figure, A1 is a unity gain difference amplifier. M1 and M2 are analog multiplier blocks with the input-output characteristic given by (3). A2 is a linear amplifier with a gain factor of 4. One of the inputs to M2 is the DC voltage equal to $\Lambda$.

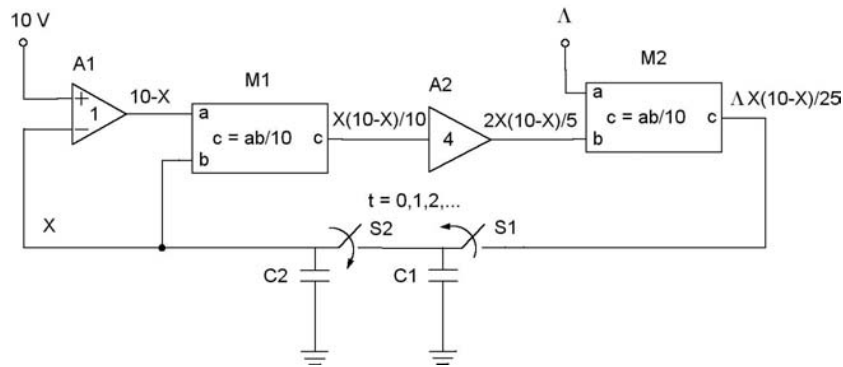

Figure 2. Schematic Block Diagram

Each of the two switch-capacitor pairs constitutes a sample-and-hold block. The switches S1 & S2 are operated in such a manner that when one of them is closed, the other is simultaneously opened. While the latest value of $X$ is being sampled by closing S1, S2 remains open and the previous value of $X$ remains on C2. After the latest value is sampled and *C1* charges to this value, *S1* is opened and *S2* is closed. Then, *C2* charges to the latest value and the next value of $X$ is computed and is available for sampling. Together, the two sample-and-hold blocks are responsible for the discrete-time nature of the system. The circuit can be called a special-purpose discrete-time analog computer designed to solve (5).

## 4. Implementation

The implementation of the scheme of figure 2 was carried out using analog ICs and components. *A1* and *A2* were implemented using the general-purpose dual operational amplifiers LM1458 from National Semiconductor. Texas Instruments / Burr Brown precision analog multiplier ICs MPY634 were used to implement *M1* and *M2* (Texas Instruments 2004). Each of the two sample-and-hold blocks was implemented using a National Semiconductor's sample-and-hold IC LF398 (National Semiconductor Corp. 2000). The trigger of the first sample-and-hold circuit was a TTL compatible square-wave generated using a 555-timer based astable multivibrator (not shown in the circuit). The second sample-and-hold was triggered using a logical inverse of this trigger. The inversion was carried out using the universal NPN bipolar junction transistor 2N3904 configured as a common-emitter switch. The collector voltage of the transistor switch becomes a square wave of 0 V and +15 V. A resistive divider of three $2.2\ \mathrm{k\Omega}$ resistors was used to convert this voltage into a TTL compatible trigger waveform. The fixed voltage of 10 V required for the difference amplifier *A1* to implement $10-X$ was obtained from the +15 V power supply by using another resistive divider. The 0-10 V voltage required as $\Lambda$ at the input of *M2* was obtained directly from a variable DC voltage source. All ICs used were Dual Inline Packages (DIP). The complete circuit diagram is shown in figure 3.

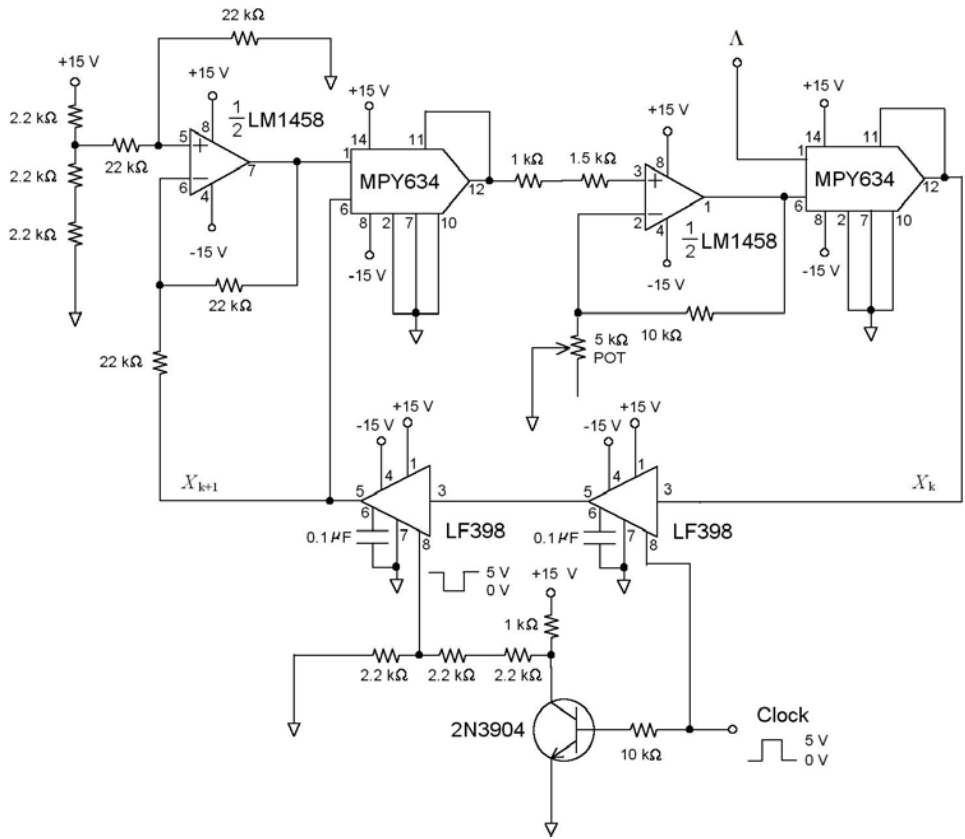

Figure 3. Circuit Diagram

$0.1\,\mu F$ Polystyrene capacitors were used as hold-capacitors for the sample-and-hold ICs. The gain of the amplifier *A2* was kept variable by providing a variable resistor (5 kΩ potentiometer) between its inverting terminal and ground. A variable gain for this amplifier helps compensate for the overall gain variation with clock frequency. The variable resistor needs to be precisely adjusted before the circuit can be ready for operation and every time the clock frequency is changed.

**5. Circuit Performance**

The behavior of the circuit of figure 3 was compared with that of computer simulations of (5). For this purpose, the bifurcation diagram of the experimental dynamical system was obtained by plotting $\Lambda$ against the long-term value of $X$ as observed at the output of the second sample-and-hold circuit (Marked $X_{k+1}$ in figure 3). An analog persistence storage oscilloscope was connected in the x-y mode for this purpose. $\Lambda$ was manually varied from 0 V to 10 V and was connected to the x-input of the oscilloscope. The output of the second sample-and-hold was connected to the y-input. The circuit was clocked at 6.5 kHz. $\Lambda$ was initially adjusted to 10 V ($\lambda = 4$) and the variable resistor was adjusted so that $X$ occupied the full interval of $(0,10)$ i.e. $x$ occupied $(0,1)$. Then $\Lambda$ was varied to obtain the bifurcation diagram. The bifurcation diagrams obtained by experiment and simulation are shown in figures 4 and 5 respectively

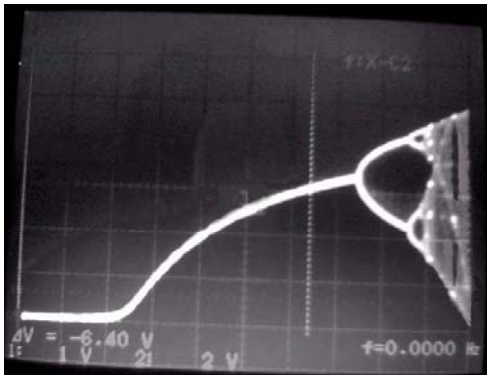

Figure 4. Bifurcation Diagram – Experiment

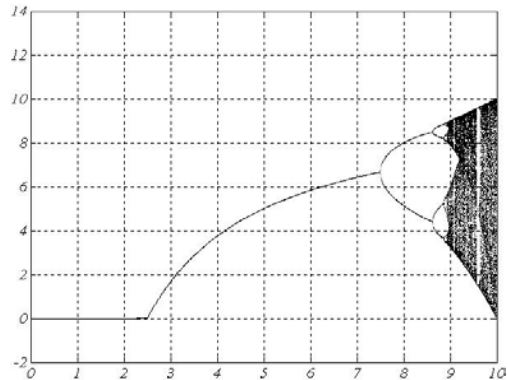

Figure 5. Bifurcation Diagram – Simulation

The portion of the diagram after the first period-doubling bifurcation was magnified by subtracting approximately 7.5 V from $\Lambda$. The resulting plot is shown in figure 6. The simulated plot of the same region of the map is shown in figure 7.

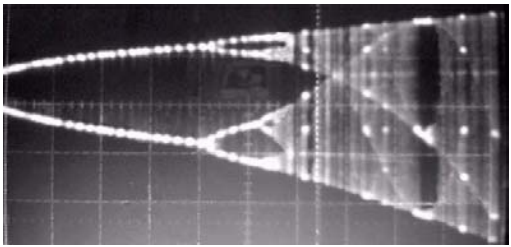

Figure 6. Magnified Bifurcation Diagram - Experiment

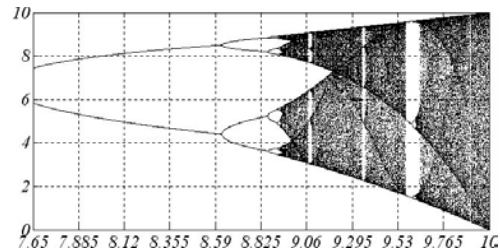

Figure 7. Magnified Bifurcation Diagram - Simulation

In figures 4 and 6, fixed points, periodic oscillations, period-doubling cascade and chaos can clearly be seen. The intermittent periodic windows are visible in the form of bright spots within chaotic regimes, the number of spots being equal to the periodicity of the orbit. In the simulated plots, the chaotic regimes are seen as darkened areas with windows of periodicity within them. From the figures, it can be seen that the circuit exhibits the entire range of behaviors of the logistic map.

Experimental measurements of $\lambda$ were carried out at various key points in the bifurcation plot such as bifurcations and intermittent periodic windows. The measured values along with those predicted by numerical simulation are shown in table 1. It can be seen that the error was found to be around 1%.

| Event | Experimental | | Theoretical | Error |
|---|---|---|---|---|
| | $\Lambda$ | $\lambda = (0.4\,\Lambda)$ | $\lambda$ | % in $\lambda$ |
| 1st Period-Doubling Bifurcation | 7.38 | 2.96 | 3 | 1.3 % |
| 2nd Period-Doubling Bifurcation | 8.52 | 3.41 | 3.44 | 0.9 % |
| 3rd Period-Doubling Bifurcation | 8.77 | 3.51 | 3.54 | 0.9 % |
| Window with Period-6 Orbit | 9.00 | 3.60 | 3.63 | 0.8 % |
| Window with Period-5 Orbit | 9.33 | 3.73 | 3.74 | 0.3 % |
| Window with Period-3 Orbit | 9.56 | 3.82 | 3.84 | 1.0 % |

Table 1. Values of $\lambda$ at some key events of the bifurcation diagram

A portion of a chaotic orbit is shown plotted against time in figure 8. Since the chaotic orbit is non-periodic, a digital storage oscilloscope was used for this purpose.

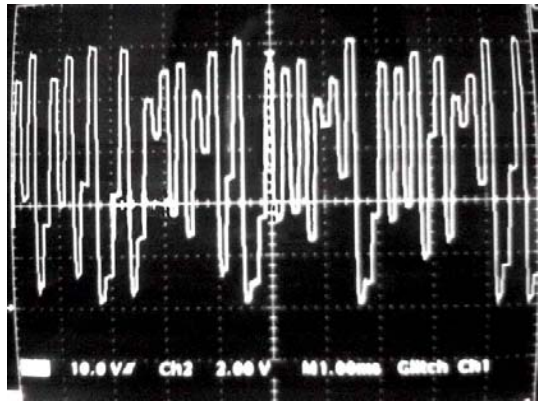

Figure 8. Snapshot of a Chaotic Orbit

## 6. Potential Applications
In this section, an attempt is made to briefly discuss a few of the possible applications of the circuit of figure 3 and other such circuits.

### 6.1 Random Number Generation
Random numbers generated by computer algorithms are only pseudo-random. They generally require an initial *seed* and two identical programs provided with the same seed generate identical random number sequences. Actual random number generators are in demand for many applications, one of which is seed generation for pseudorandomness algorithms. Circuits such as the proposed one may be used as inexpensive and compact random number generators. The circuit may be clocked at a high frequency and the state of the system may be read once in, say, ten thousand clock cycles. Because of extreme sensitivity to initial conditions, the resulting sequence will be unpredictable. Two such circuits provided with practically identical conditions would not generate identical sequences.

### *6.2 Frequency-Hopping*

Considering the fact that a chaotic orbit is unpredictable in the long run, it is natural to see the application as the basis of a frequency hopper for radar or communication applications. The frequency of the transmitter can be varied as a function of the state of the chaotic system. Since the transmitter and receiver of a radar system are co-located, a single chaos generator can be used for both. For a communication system, however, it is necessary to synchronize the chaos generators at both ends so that the transmitter and the receiver hop as per the same chaotic orbit. The problem of chaos synchronization is still under investigation (Pethel et al 2002). The use of numerically computed chaotic sequences for Frequency Hopping Code Division Multiple Access (FH/CDMA) has been previously proposed (Cong & Songgeng 2002). However, as seen in § 2, computation algorithms are necessarily finite state machines that give rise to periodic orbits. Such sequences are therefore predictable whereas chaotic hardware based frequency hoppers rely on truly chaotic non-periodic sequences.

### *6.3 Ranging*

The non-recurring nature of chaotic signals eliminates range ambiguities present in ranging systems employing recurring sequences. Further, the inherently wideband nature of the chaotic signal affords better range resolution. The sequences also provide resistance against jamming in a hostile environment. In a military radar application, this feature can even be combined with chaotic frequency hopping to introduce strong jammer / interceptor avoidance capabilities.

### *6.4 Spread-Spectrum Communications*

The proposed hardware can be used with any of the chaos-based spread-spectrum communication schemes such as Chaos Shift-Keying, Differential Chaos Shift-Keying (DCSK) and Modified DCSK (M-DCSK) for single or multi-user communications (Heidari-Bateni & McGillem 1994, Galias & Maggio 2001, Mandal 2002, Mandal & Banerjee 2003).

### 7. Conclusion

An electronic circuit has been designed, implemented and tested to accurately realize the logistic difference equation. The circuit has been found to replicate the whole range of behaviors of the logistic map. It has also been shown to be reasonably accurate. Though only the logistic equation has been realized in the current work, it is straightforward to implement other systems on similar lines. The techniques employed are simple and the approach can be extended to realize other types of maps such as piecewise linear or piecewise smooth maps. Similar realizations of higher-dimensional maps such as the two-dimensional Henon map are also feasible. Such realizations have many potential applications, a few of which have been briefly discussed. The ones mentioned in this paper, however, are in no way exhaustive. As with most phenomena in non-linear dynamics, there are bound to be inter-disciplinary applications. Such circuits are also useful research tools - laboratory models of mathematical systems with complex behavior.

The author is grateful to Dr V K Saraswat for encouragement. He is also thankful to A S Sarma, and B Jaya for their support and encouragement. The experimental work was made possible by help from several colleagues, especially S S Reddy Kumar, D Venu Gopal and H Madhukar.

**List of Symbols**

| Symbol | Description |
|---|---|
| $a, b$ | Input voltages of a practical analog multiplier |
| $c$ | Output voltage of a practical analog multiplier |
| $f$ | Iterative map describing a discrete-time dynamical system |
| $k$ | Discrete time |
| $n$ | Dimensionality of state-space |
| $\Re$ | The set of real numbers |
| $x$ | State-variable of a discrete-time dynamical system |
| $X$ | $10 x$ |
| $\lambda$ | Control (bifurcation) parameter of logistic equation |
| $\Lambda$ | $2.5 \lambda$ |